\documentclass[final]{IEEEtran}
\pdfoutput=1
\usepackage{amsthm,amssymb,graphicx,multirow,amsmath,color,amsfonts}
\usepackage[update,prepend]{epstopdf}
\usepackage[noadjust]{cite}
\usepackage[latin1]{inputenc}
\usepackage{tikz}
\usepackage{bbm} 
\usepackage{pdfpages}
\usepackage{tabulary}
\usepackage{multirow}
\usepackage{comment}

\newtheorem{thm}{Theorem}
\newtheorem{lemm}{Lemma}
\newtheorem{cor}{Corollary}

\include{notation}
\allowdisplaybreaks 

\begin{document}
\graphicspath{{./Figures/}}
\title{
Closed-Form Coverage Probability for Downlink Poisson Network With Double Shadowed Fading}
\author{
Jingrui~Chen and Chaowei~Yuan
\thanks{Jingrui Chen and Chaowei Yuan are with the School of Information and Communication Engineering, Beijing University of Posts and Telecommunications, Beijing, 100876, China, (e-mail: xiaojingyun604604@163.com; yuancw2000@bupt.edu.cn). 
Last updated: \today.}
\vspace{-10mm}
}
\maketitle
\begin{abstract}
Performances of cellular networks over $\kappa$-$\mu$ shadowed fading with long-term shadowing has been studied in the existing literature. However, the impact of $\kappa$-$\mu$ shadowed fading with instantaneous shadowing on performances of cellular networks is unknown. Therefore, this letter analyzes the downlink coverage probability of a Poisson network with double shadowed fading which is composed of a large-scale fading of lognormal distribution and $\kappa$-$\mu$ shadowed fading with integer fading parameters. The closest base station association rule without shadowing is considered. For analytical tractability, the double shadowed fading is approximated as a weighted sum of $\kappa$-$\mu$ shadowed distributions based on the Gaussian-Hermit quadrature. As a main theoretical result, a closed-form expression for the downlink coverage probability of a Poisson network under double shadowed fading for the desired signal and arbitrary fading for the interfering signals is successfully derived. Numerical simulations reveal that the double shadowed fading provides a pessimistic coverage on a Poisson network compared with the long-term shadowing which is incorporated into cell selection.

\end{abstract}
\vspace{-2mm}
\section{Introduction} \label{sec:intro}
Stochastic geometry as a useful tool has employed to the performance analysis of heterogeneous cellular networks (HetNets), where the irregular locations of users and base stations (BSs) are modeled by Point processes~\cite{Elsawy2013cst,andrews2016online}. The single-tier cellular network is developed in~\cite{andrews2011tc} and then extended to a multi-tier HetNet~\cite{dhillon2012jsac}. To meet the needs of various fading scenarios in future communications, a versatile fading model, i.e. $\kappa$-$\mu$ shadowed fading is proposed in~\cite{paris2014tvt}. However, it is intractable to the performance analysis of HetNets over this fading model due to the existence of the confluent hypergeometric function (CHF). To overcome this problem, the moment matching method with Gamma distribution is utilized in~\cite{kumar2015wcl} and the truncated series form of the CHF is used in~\cite{parthasarathy2017wcl}.

In order to accommodate the requirements of various fading scenarios in fifth-generation communications, a double shadowed fading model is first proposed in~\cite{chun2017twc}, where the double shadowed fading is composed of a large-scale fading and $\kappa$-$\mu$ shadowed fading. Similar to~\cite{dhillon2014wcl,Saha2017twc}, the large-scale fading is incorporated into cell association, which can characterize the impact of long-term shadowing on the cell selection. As a result, the effect of $\kappa$-$\mu$ shadowed fading with long-term shadowing on performances of cellular networks is studied.

In this paper, we distinguish from the previous related works~\cite{chun2017twc}-\cite{Saha2017twc} by studying the impact of instantaneous shadowing on the coverage probability of a Poisson network. Specifically, we consider a double shadowed fading which is composed of a lognormal distribution and $\kappa$-$\mu$ shadowed fading with integer fading parameters~\cite{martinez2017tvt}. The nearest BS selection without shadowing is also considered.

There are three contributions in this paper. Firstly, an accurate and tractable expression for the double shadowed fading is obtained according to the Gaussian-Hermit quadrature (GHQ)~\cite{abramowitz1972book}. Secondly, a closed-form expression for the downlink coverage probability of a Poisson network is successfully derived, assuming that the desired signal experiences double shadowed fading and the interfering signals experience arbitrary fading. The resultant expression avoids complex integral evaluations or high order differential calculations. Meanwhile, it is generic due to the application of versatile model of double shadowed fading in the desired link and the assumption of an arbitrary fading distribution for all interfering links. Most importantly, the theoretical analysis of coverage probability with respect to the interfering fading is avoided under the independent assumptions of fading models. Thirdly, compared with the long-term shadowed environment, numerical simulations show that the double shadowed fading offers a smaller coverage to a Poisson network.

\vspace{-0.5mm}
\section{System Model}
Consider a downlink Poisson network where the locations of BSs are modeled by a homogeneous Poisson point process ${\Phi _{\bf{B}}}$ with density ${\lambda}$. Assuming that users are uniformly distributed in the network. The received power of a user at origin (termed as a typical user) from a BS at ${\bf{x}} \in {\Phi _{\rm{B}}}$ is modeled as $P\left( {\bf{x}} \right) = P{H_{\bf{x}}}{\left\| {\bf{x}} \right\|^{ - \alpha }}$, where $P$ is the transmit power, $\alpha$ is the path loss exponent, and $H_{\bf{x}}$ is the independent fading gain with unit power. Consider the nearest BS association policy, thus the location of serving BS is ${\bf{x}}* = \arg {\max _{{\bf{x}} \in {\Phi _{\rm{B}}}}}P{\left\| {\bf{x}} \right\|^{ - \alpha }}$. According to~\cite{andrews2011tc}, the probability density function (PDF) of serving distance is ${f_R}\left( r \right) = 2\pi \lambda r\exp \left( { - \pi \lambda {r^2}} \right)$. Assuming that the channel power of the desired signal ${H_{{\bf{x}}*}}$ follows a double shadowed distribution, i.e. ${H_{{\bf{x}}*}} = {H_\chi }\cdot{H_S}$, where the independent random variables ${H_\chi }$ follows a lognormal distribution with a standard deviation (SD) ${\tilde \sigma _S}$ (${\tilde \sigma _S} = {{{\sigma _S}\ln \left( {10} \right)} \mathord{\left/
 {\vphantom {{{\sigma _S}\ln \left( {10} \right)} {10}}} \right.
 \kern-\nulldelimiterspace} {10}}$) and $H_S$ follows a $\kappa$-$\mu$ shadowed distribution with integer fading parameters~\cite{martinez2017tvt}. For notational simplicity, we denote the distributions ${H_\chi }$ and $H_S$ as ${H_\chi } \sim \ln N\left( {0,{{\tilde \sigma }_S}} \right)$ and ${H_S} \sim S\left( {\kappa ,\mu ,m;{h_S}} \right)$ respectively. According to~\cite[Eq.~(12), Eq.~(13)]{martinez2017tvt}, the PDF and complementary cumulative distribution function (CCDF) of ${H_S}$ are
\begin{align}\label{1}
{f_{{H_S}}}\left( {\kappa ,\mu ,m;{h_S}} \right) = \sum\limits_{i = 0}^M {\frac{{{C_i}h_S^{{m_i} - 1}}}{{\Gamma \left( {{m_i}} \right){\Omega _i}^{{m_i}}}}\exp \left( { - \frac{{{h_S}}}{{{\Omega _i}}}} \right)},
\end{align}
\begin{align}\label{2}
{\bar F_{{H_S}}}\left( {\kappa ,\mu ,m;{h_S}} \right) = \sum\limits_{i = 0}^M {\sum\limits_{j = 0}^{{m_i} - 1} {\frac{{{C_i}{{\left( {{{{h_S}} \mathord{\left/
 {\vphantom {{{h_S}} {{\Omega _i}}}} \right.
 \kern-\nulldelimiterspace} {{\Omega _i}}}} \right)}^j}}}{{\Gamma \left( {j + 1} \right)\exp \left( {{{{h_S}} \mathord{\left/
 {\vphantom {{{h_S}} {{\Omega _i}}}} \right.
 \kern-\nulldelimiterspace} {{\Omega _i}}}} \right)}}} },
\end{align}
where the coefficients $M$, ${m_i}$, ${\Omega _i}$ and ${C_i}$ are depended on the fading parameters $\kappa$, $\mu$ and $m$. Let ${\omega _1} = \frac{m}{{\kappa \mu  + m}}$, ${\omega _2} = \frac{{\kappa \mu }}{{\kappa \mu  + m}}$ and ${\omega _3} = \frac{1}{{\mu \left( {\kappa  + 1} \right)}}$. According to~\cite[Table I]{martinez2017tvt}, if $\mu>m$, then $M=\mu$, ${m_i} = \mu  - m - i + 1$, ${\Omega _i} = {\omega _3}$ and ${C_i} = \left( {\begin{array}{*{20}{c}}
{m + i - 2}\\
{i - 1}
\end{array}} \right){\left( { - {\omega _1}} \right)^m}\omega _2^{ 1- m - i }$ for $i = 1,2, \ldots ,\mu  - m$, and ${m_i} = \mu  - i + 1$, ${\Omega _i} = \frac{{{\omega _3}}}{{{\omega _1}}}$ and ${C_i} = \left( {\begin{array}{*{20}{c}}
{i - 2}\\
{i - \mu  + m - 1}
\end{array}} \right){\left( { - {\omega _1}} \right)^{i - \mu  + m - 1}}\omega _2^{1 - i}$ for $i = \mu  - m + 1, \ldots ,\mu$, otherwise $M = m - \mu$, ${m_i} = m - i$, ${\Omega _i} = \frac{{{\omega _3}}}{{{\omega _1}}}$ and ${C_i} = \left( {\begin{array}{*{20}{c}}
{m - \mu }\\
i
\end{array}} \right)\omega _1^i\omega _2^{m - \mu  - i}$. Assuming that the channel powers of the interfering signals $\left\{ {{H_{\bf{x}}}} \right\}$ are i.i.d., where ${\bf{x}} \in {\Phi _{\rm{B}}}\backslash {\bf{x}}*$. In order to obtain a generic result, we consider ${H_{\bf{x}}}$ follows an arbitrary distribution. The orthogonal multiple access technology~\cite{andrews2011tc} is employed at each cell to eliminate intra-cell interference. Thus, by considering the interference-limited (noise is ignored) network, the signal-to-interference ratio (SIR) of a typical user can be written as ${\rm{SIR}}\left( r \right) = \frac{{P{H_{{\bf{x}}*}}{r^{ - \alpha }}}}{I}$, where the aggregate interference power $I = \sum\nolimits_{{\bf{x}} \in {\Phi _{\rm{B}}}\backslash {\bf{x}}*} {P{H_{\bf{x}}}{{\left\| {\bf{x}} \right\|}^{ - \alpha }}}$.

\vspace{-1.5mm}
\section{Double shadowed fading}
\begin{figure}
\centering
\includegraphics[width=0.65\columnwidth]{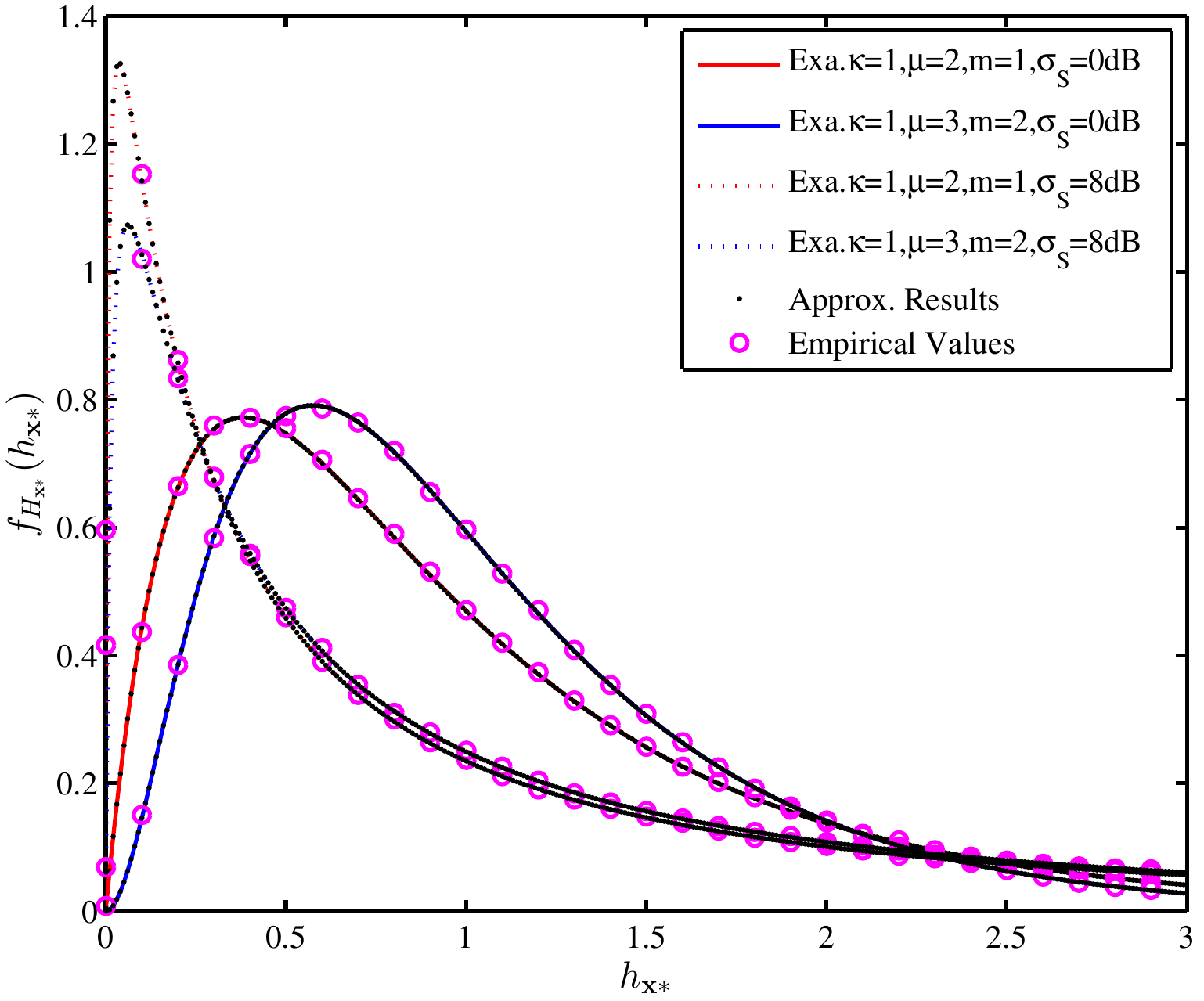}
\caption{PDF of double shadowed fading for $n_{\rm{GHQ}}=32$.}
\label{fig:1}
\end{figure}

In order to study the impact of $\kappa$-$\mu$ shadowed fading with instantaneous shadowing on performances of cellular networks, the exact PDF of the double shadowed distribution described in section II is given bellow.
\begin{lemm}\label{Lemma:1}
Given two mutually independent random variables ${H_\chi }$ and $H_S$, where ${H_\chi } \sim \ln N\left( {0,{{\tilde \sigma }_S}} \right)$ and ${H_S} \sim S\left( {\kappa ,\mu ,m;{h_S}} \right)$, the exact PDF of the double shadowed fading ${H_{{\bf{x}}*}} = {H_\chi }\cdot{H_S}$ is
 \begin{align}\label{3}
{f_{{H_{{\bf{x}}*}}}}\left( {{h_{{\bf{x}}*}}} \right) = \sum\limits_{i = 0}^M {\frac{{\int_0^\infty  {\frac{{h_S^{{m_i} - 1}}}{{\exp \left( {\frac{{{h_S}}}{{{\Omega _i}}} + \frac{{{{\left( {\ln {h_{{\bf{x}}*}} - \ln {h_S}} \right)}^2}}}{{2\tilde \sigma _S^2}}} \right)}}d{h_S}} }}{{\sqrt {2\pi } {{\tilde \sigma }_{\rm{S}}}\Gamma \left( {{m_i}} \right)C_i^{ - 1}\Omega _i^{{m_i}}{h_{{\bf{x}}*}}}}},
 \end{align}
where the coefficients $M$, ${m_i}$, ${\Omega _i}$ and ${C_i}$ are given by~(\ref{1}).
\end{lemm}
\begin{IEEEproof}
Based on the density function of product of two random variables~\cite{papoulis1991book}, the PDF ${H_{{\bf{x}}*}} = {H_\chi }\cdot{H_S}$ is
\begin{align}\label{4}
{f_{{H_{{\bf{x}}*}}}}\left( {{h_{{\bf{x}}*}}} \right) = \int_0^\infty  {{f_{{H_S}}}\left( {{h_S}} \right){f_{{H_\chi }}}\left( {\frac{{{h_{{\bf{x}}*}}}}{{{h_S}}}} \right)\frac{1}{{\left| {{h_S}} \right|}}d{h_S}}.
\end{align}
Substituting the distribution of ${H_\chi } \sim \ln N\left( {0,{{\tilde \sigma }_S}} \right)$ and~(\ref{1}) into~(\ref{4}) yields~(\ref{3}). This completes the proof.
\end{IEEEproof}
It has been known that the $\kappa$-$\mu$ shadowed fading contains majority of the fading models proposed in the literature as special cases~\cite{chun2017twc}, including Rayleigh, Rician, Nakagami-$m$, Nakagami-$q$, One-sided Gaussian, $\kappa$-$\mu$, $\eta$-$\mu $, etc. Therefore, Lemma~\ref{Lemma:1} encompasses various composite fading models and shows the instantaneous property of large scale fading of lognormal distribution. However, it is intractable to the performance analysis of cellular networks due to an integral existed in~(\ref{3}). To conquer this problem, the GHQ rule~\cite[Eq.~(25.4.46)]{abramowitz1972book} is employed and the derived expression is as follows.
\begin{lemm}\label{Lemma:2}
The double shadowed distribution described in Lemma~\ref{Lemma:1} can be expressed as
\begin{align}\label{5}
{f_{{H_{{\bf{x}}*}}}}\left( {{h_{{\bf{x}}*}}} \right) = \sum\limits_{l = 1}^{{n_{{\rm{GHQ}}}}} {{a_l}{{f}_{{H_S}}}\left( {\kappa ,\mu ,m;{{{h_{{\bf{x}}*}}} \mathord{\left/
 {\vphantom {{{h_{{\bf{x}}*}}} {{b_l}}}} \right.
 \kern-\nulldelimiterspace} {{b_l}}}} \right)},
\end{align}
where ${a_l} = {w_l}/\sum\limits_{{l_1} = 1}^{{n_{{\rm{GHQ}}}}} {{w_{{l_1}}}} $, ${b_l} = \exp \left( {\sqrt 2 {{\tilde \sigma }_{\rm{S}}}{t_{{l_1}}}} \right)$, ${w_{{l_1}}}$ and ${t_{{l_1}}}$ are the weights and abscissas of the ${l_1}$-th order Hermite polynomial respectively, and ${f_{{H_S}}}\left( \cdot \right)$ is given by~(\ref{1}).
\end{lemm}
\begin{IEEEproof}
The density function of product of two random variables~\cite{papoulis1991book} ${H_{{\bf{x}}*}} = {H_\chi }\cdot{H_S}$ can be rewritten as
\begin{align}\label{6}
{f_{{H_{{\bf{x}}*}}}}\left( {{h_{{\bf{x}}*}}} \right) = \int_0^\infty  {{f_{{H_\chi }}}\left( {{h_\chi }} \right){f_{{H_S}}}\left( {\frac{{{h_{{\bf{x}}*}}}}{{{h_\chi }}}} \right)\frac{1}{{\left| {{h_\chi }} \right|}}d{h_\chi }}.
\end{align}
Substituting the distribution of ${H_\chi } \sim \ln N\left( {0,{{\tilde \sigma }_S}} \right)$ and~(\ref{1}) into~(\ref{6}), we can obtain
\begin{align}\label{7}
{f_{{H_{{\bf{x}}*}}}}\left( {{h_{{\bf{x}}*}}} \right) = \sum\limits_{i = 0}^M {\frac{{\int_0^\infty  {\frac{{\exp \left( { - \frac{{{h_{{\bf{x}}*}}}}{{{\Omega _i}{h_\chi }}} - \frac{{{{\left( {\ln {h_\chi }} \right)}^2}}}{{2\tilde \sigma _S^2}}} \right)}}{{h_\chi ^{{m_i} + 1}}}d{h_\chi }} }}{{\sqrt {2\pi } {{\tilde \sigma }_{\rm{S}}}\Gamma \left( {{m_i}} \right)C_i^{ - 1}\Omega _i^{{m_i}}h_{{\bf{x}}*}^{1 - {m_i}}}}}.
\end{align}
Using the change of variable $t = \frac{{\ln {h_\chi }}}{{\sqrt 2 {{\tilde \sigma }_{\rm{S}}}}}$ in~(\ref{7}), applying the GHQ rule and combing with the identity $\sqrt \pi   = \sum\nolimits_{{l_1} = 1}^{{n_{{\rm{GHQ}}}}} {{w_{{l_1}}}}$ yields~(\ref{5}). This completes the proof.
\end{IEEEproof}
As illustrated in Fig.~\ref{fig:1}, the approximated expression provided in~(\ref{5}) exactly matches the exact expression in Lemma~\ref{Lemma:1}. Most importantly, the ${f_{{H_{{\bf{x}}*}}}}\left( {{h_{{\bf{x}}*}}} \right)$ in Lemma~\ref{Lemma:2} is expressed as a weighted sum of $\kappa$-$\mu$ shadowed PDFs ${f_{{H_S}}}\left( \cdot \right)$, which simplifies the analytical expressions significantly in special cases. When the lognormal distribution is ignored,~(\ref{5}) is reduced to~(\ref{1}) as ${\sigma _{\rm{S}}} = 0$dB. Another important expression is the CCDF of ${H_{{\bf{x}}*}}$, which is directly utilized for the coverage analysis of a downlink Poisson network. Combing with~(\ref{2}) and Lemma~\ref{Lemma:2}, we can obtain the CCDF of ${H_{{\bf{x}}*}}$ conveniently.
\begin{cor}\label{Corollary:1}
The CCDF of the double shadowed fading is
\begin{align}\label{8}
{\bar F_{{H_{{\bf{x}}*}}}}\left( {{h_{{\bf{x}}*}}} \right) = \sum\limits_{l = 1}^{{n_{{\rm{GHQ}}}}} {{a_l}{{\bar F}_{{H_S}}}\left( {\kappa ,\mu ,m;{{{h_{{\bf{x}}*}}} \mathord{\left/
 {\vphantom {{{h_{{\bf{x}}*}}} {{b_l}}}} \right.
 \kern-\nulldelimiterspace} {{b_l}}}} \right)},
\end{align}
where the coefficients ${a_l}$ and ${b_l}$ are given by Lemma~\ref{Lemma:2} and ${\bar F_{{H_S}}}\left( \cdot \right)$ is given by~(\ref{2}).
\end{cor}

\vspace{-2mm}
\section{Coverage Probability}
The coverage probability in a downlink Poisson network is defined as the probability that the received power of a typical user is bigger than the SIR threshold $\theta$~\cite{andrews2011tc}. Consider the nearest BS association policy, the closed-form expression for the downlink coverage probability in a Poisson network is obtained in the next Theorem.
\begin{thm}\label{Theorem:1}
When the desired signal experiences the double shadowed fading and the interfering signals experience arbitrary fading, the coverage probability of a typical user is
\begin{align}\label{9}
{p_{\rm{c}}}\left( \theta  \right) = \sum\limits_{i = 0}^M {\sum\limits_{j = 0}^{{m_i} - 1} {\sum\limits_{l = 1}^{{n_{{\rm{GHQ}}}}} {\sum\limits_{\left( {{N_1}, \cdots ,{N_j}} \right) \in {\mathbb{T}_j}} {\frac{{{a_l}{C_i}{A_j}\Gamma \left( {{B_j}} \right)}}{{{{\left( { - 1} \right)}^j}G_i^{{B_j}}}}} } } },
\end{align}
where the coefficients $M$, ${m_i}$, ${\Omega _i}$, ${C_i}$, ${a_l}$ and ${b_l}$ are given by Corollary~\ref{Corollary:1}, ${\mathbb{T}_j} = \left\{ {\left. {\left( {{N_1}, \cdots ,{N_j}} \right) \in {{\left( {\mathbb{N} \cup \left\{ 0 \right\}} \right)}^j}} \right|\sum\limits_{q = 1}^j {{N_q}q}  = j,\sum\limits_{q = 1}^j {{N_q}}  = q} \right\}$, ${A_j} = \prod\limits_{q = 1}^j {\frac{1}{{\Gamma \left( {{N_q} + 1} \right)}}{{\left( {\frac{{2{{\left( { - 1} \right)}^q}{E_{{h_q}}}}}{{\alpha \Gamma \left( {q + 1} \right)}}{{\left( {\frac{\theta }{{{\Omega _i}{b_l}}}} \right)}^{{2 \mathord{\left/
 {\vphantom {2 \alpha }} \right.
 \kern-\nulldelimiterspace} \alpha }}}} \right)}^{{N_q}}}}$, ${B_j} = 1 + \sum\limits_{q = 1}^j {{N_q}}$, ${G_i} = 1 + {\left( {{\theta  \mathord{\left/
 {\vphantom {\theta  {{\Omega _i}{b_l}}}} \right.
 \kern-\nulldelimiterspace} {{\Omega _i}{b_l}}}} \right)^{{2 \mathord{\left/
 {\vphantom {2 \alpha }} \right.
 \kern-\nulldelimiterspace} \alpha }}}{E_{{h_0}}}$, the values of the expectations with respect to the interfering fading distributions are ${E_{{h_0}}} = {\mathbb{E}_{{H_{\bf{x}}}}}\left[ {\int_{{{\left( {{{{\Omega _i}{b_l}} \mathord{\left/
 {\vphantom {{{\Omega _i}{b_l}} \theta }} \right.
 \kern-\nulldelimiterspace} \theta }} \right)}^{{2 \mathord{\left/
 {\vphantom {2 \alpha }} \right.
 \kern-\nulldelimiterspace} \alpha }}}}^\infty  {\left( {1 - \exp \left( { - {h_{\bf{x}}}{t^{{{ - \alpha } \mathord{\left/
 {\vphantom {{ - \alpha } 2}} \right.
 \kern-\nulldelimiterspace} 2}}}} \right)} \right)dt} } \right]$ and ${E_{{h_q}}} = {\mathbb{E}_{{H_{\bf{x}}}}}\left[ {h_{\bf{x}}^{{2 \mathord{\left/
 {\vphantom {2 \alpha }} \right.
 \kern-\nulldelimiterspace} \alpha }}\gamma \left( {q - {2 \mathord{\left/
 {\vphantom {2 \alpha }} \right.
 \kern-\nulldelimiterspace} \alpha },\frac{{\theta {h_{\bf{x}}}}}{{{\Omega _i}{b_l}}}} \right)} \right]$, and the functions $\Gamma \left(\cdot\right)$ and $\gamma \left( {a,x} \right)$ are the Gamma and lower incomplete Gamma function~\cite{gradshteyn2007book} respectively.
\end{thm}

\begin{IEEEproof}
Substituting the expression of SIR into the definition of the coverage probability, we have
\begin{align}\label{10}
{p_{\rm{c}}}\left( {\theta ,r} \right) = \mathbb{P}\left( {{h_{{\bf{x}}*}} > \frac{{\theta {I_1}{r^\alpha }}}{P}} \right).
\end{align}
Since the ${H_{{\bf{x}}*}}$ follows the double shadowed fading, combined by~(\ref{8}) in Corollary~\ref{Corollary:1}, we obtain
\begin{align}\label{11}
{p_{\rm{c}}}\left( {\theta ,r} \right) = \sum\limits_{i = 0}^M {\sum\limits_{j = 0}^{{m_i} - 1} {\sum\limits_{l = 1}^{{n_{{\rm{GHQ}}}}} {\frac{{{C_i}{a_l}{{\left( {\frac{{\theta I{r^\alpha }}}{{P{\Omega _i}{b_l}}}} \right)}^j}}}{{\Gamma \left( {j + 1} \right)\exp \left( {\frac{{\theta I{r^\alpha }}}{{P{\Omega _i}{b_l}}}} \right)}}} } }.
\end{align}
Due to the fact that $\exp \left( { - \alpha } \right) = {\left( { - \alpha } \right)^{ - j}}{\left. {\exp {{\left( { - \alpha z} \right)}^{\left( j \right)}}} \right|_{z = 1}}$, where ${g^{\left( j \right)}}\left( z \right)$ is the $j$-th differential of $z$,~(\ref{11}) can be written as
\begin{align}\label{12}
{p_{\rm{c}}}\left( {\theta ,r} \right) = \sum\limits_{i = 0}^M {\sum\limits_{j = 0}^{{m_i} - 1} {\sum\limits_{l = 1}^{{n_{{\rm{GHQ}}}}} {\frac{{{{\left. {\frac{{{\partial ^j}}}{{\partial {z^j}}}\mathbb{E}\left[ {\exp \left( { - \frac{{zsI}}{{{\Omega _i}{b_l}}}} \right)} \right]} \right|}_{z = 1}}}}{{a_l^{ - 1}C_i^{ - 1}{{\left( { - 1} \right)}^j}\Gamma \left( {j + 1} \right)}}} } },
\end{align}
where $s = {{\theta {r^\alpha }} \mathord{\left/
 {\vphantom {{\theta {r^\alpha }} P}} \right.
 \kern-\nulldelimiterspace} P}$. Since ${H_{\bf{x}}}$ has to be identically distributed as well as for all interfering BSs, substituting $I = \sum\nolimits_{{\bf{x}} \in {\Phi _{\bf{B}}}\backslash \left\{ {{\bf{x}}*} \right\}} {P{H_{\bf{x}}}{{\left\| {\bf{x}} \right\|}^{ - \alpha }}}$ into~(\ref{12}) and using the result of the conditional generating function of PPP~\cite{andrews2011tc}, we get
\begin{align}\label{13}
\begin{array}{l}
{p_{\rm{c}}}\left( {\theta ,r} \right) = \sum\limits_{i = 0}^M {\sum\limits_{j = 0}^{{m_i} - 1} {\sum\limits_{l = 1}^{{n_{{\rm{GHQ}}}}} {\frac{{{C_i}{a_l}{{\left( { - 1} \right)}^j}}}{{\Gamma \left( {j + 1} \right)}}\frac{{{\partial ^j}}}{{\partial {z^j}}}\exp \left( { - 2\pi \lambda } \right.} } } \\
{\left. { \times \left. {{\mathbb{E}_{{H_{\bf{x}}}}}\left[ {\int_r^\infty  {\left( {1 - \exp \left( { - zL{h_{\bf{x}}}} \right)} \right)vdv} } \right]} \right)} \right|_{z = 1}},
\end{array}
\end{align}
where $L = \frac{{\theta {r^\alpha }}}{{{\Omega _i}{b_l}{v^\alpha }}}$. According to the higher order derivatives~\cite[Eq. (04.30.2)]{gradshteyn2007book}, we obtain
\begin{align}\label{14}
\begin{array}{l}
\frac{{{\partial ^j}}}{{\partial {z^j}}}\exp \left( { - 2\pi \lambda {\mathbb{E}_{{H_{\bf{x}}}}}\left[ {\int_r^\infty  {\left( {1 - \exp \left( { - zL{h_{\bf{x}}}} \right)} \right)vdv} } \right]} \right)\\
 = j!\exp \left( { - 2\pi \lambda {\mathbb{E}_{{H_{\bf{x}}}}}\left[ {\int_r^\infty  {\left( {1 - \exp \left( { - zL{h_{\bf{x}}}} \right)} \right)vdv} } \right]} \right)\\
 \times \sum\limits_{\left( {{N_1}, \cdots ,{N_j}} \right) \in {\mathbb{T}_j}} {\prod\limits_{q = 1}^j {\frac{1}{{\Gamma \left( {{N_q} + 1} \right)}}\left( {\frac{1}{{\Gamma \left( {q + 1} \right)}}} \right.\frac{{{\partial ^q}}}{{\partial {z^q}}}\left( { - 2} \right.} } \\
{\left. {\left. { \times \pi \lambda {\mathbb{E}_{{H_{\bf{x}}}}}\left[ {\int_r^\infty  {\left( {1 - \exp \left( { - zL{h_{\bf{x}}}} \right)} \right)vdv} } \right]} \right)} \right)^{{D_q}}},
\end{array}
\end{align}
where ${\mathbb{T}_j}$ is given in Theorem~\ref{Theorem:1}. Solving the derivative $\frac{{{\partial ^q}}}{{\partial {z^q}}}\left( { - 2\pi {\lambda}{\mathbb{E}_{{H_{\bf{x}}}}}\left[ {\int_r^\infty  {\left( {1 - \exp \left( { - zL{h_{\bf{x}}}} \right)} \right)vdv} } \right]} \right)$, plugging in $z=1$ and $L$, and by using the change of variable $t = {\left( {{\theta  \mathord{\left/
 {\vphantom {\theta  {{\Omega _i}{b_l}}}} \right.
 \kern-\nulldelimiterspace} {{\Omega _i}{b_l}}}} \right)^{{{ - 1} \mathord{\left/
 {\vphantom {{ - 1} \alpha }} \right.
 \kern-\nulldelimiterspace} \alpha }}}{r^{ - 1}}v$, we have
 \begin{align}\label{15}
 \begin{array}{l}
\frac{{{\partial ^q}}}{{\partial {z^q}}}\left( { - 2\pi \lambda {\mathbb{E}_{{H_{\bf{x}}}}}\left[ {\int_r^\infty  {\left( {1 - \exp \left( { - zL{h_{\bf{x}}}} \right)} \right)vdv} } \right]} \right)\\
 = \frac{{2\pi \lambda {r^2}{{\left( { - 1} \right)}^q}{E_{{h_{\bf{x}}}q}}}}{\alpha }{\left( {\frac{\theta }{{{\Omega _i}{b_l}}}} \right)^{{2 \mathord{\left/
 {\vphantom {2 \alpha }} \right.
 \kern-\nulldelimiterspace} \alpha }}},
\end{array}
\end{align}
where ${E_{{h_q}}} = {\mathbb{E}_{{H_{\bf{x}}}}}\left[ {h_{\bf{x}}^{{2 \mathord{\left/
 {\vphantom {2 \alpha }} \right.
 \kern-\nulldelimiterspace} \alpha }}\gamma \left( {q - {2 \mathord{\left/
 {\vphantom {2 \alpha }} \right.
 \kern-\nulldelimiterspace} \alpha },\frac{{\theta {h_{\bf{x}}}}}{{{\Omega _i}{b_l}}}} \right)} \right]$ . In addition, substituting $z=1$ and $L$ into ${\mathbb{E}_{{H_{\bf{x}}}}}\left[ {\int_r^\infty  {\left( {1 - \exp \left( { - zL{h_{\bf{x}}}} \right)} \right)vdv} } \right]$ and by using the change of variable $t = \frac{{\theta {r^\alpha }{h_{\bf{x}}}}}{{{\Omega _i}{b_l}{v^\alpha }}}$, we obtain
 \begin{align}\label{16}
\begin{array}{l}
\exp \left( { - 2\pi \lambda {\mathbb{E}_{{H_{\bf{x}}}}}\left[ {\int_r^\infty  {\left( {1 - \exp \left( { - zL{h_{\bf{x}}}} \right)} \right)vdv} } \right]} \right)\\
 = \exp \left( { - \pi \lambda {r^2}{{\left( {\frac{\theta }{{{\Omega _i}{b_l}}}} \right)}^{{2 \mathord{\left/
 {\vphantom {2 \alpha }} \right.
 \kern-\nulldelimiterspace} \alpha }}}{E_{{h_0}}}} \right),
\end{array}
 \end{align}
where ${E_{{h_0}}} = {\mathbb{E}_{{H_{\bf{x}}}}}\left[ {\int_{{{\left( {{{{\Omega _i}{b_l}} \mathord{\left/
 {\vphantom {{{\Omega _i}{b_l}} \theta }} \right.
 \kern-\nulldelimiterspace} \theta }} \right)}^{{2 \mathord{\left/
 {\vphantom {2 \alpha }} \right.
 \kern-\nulldelimiterspace} \alpha }}}}^\infty  {\left( {1 - \exp \left( { - {h_{\bf{x}}}{t^{{{ - \alpha } \mathord{\left/
 {\vphantom {{ - \alpha } 2}} \right.
 \kern-\nulldelimiterspace} 2}}}} \right)} \right)dt} } \right]$. Substituting~(\ref{14}),~(\ref{15}) and~(\ref{16}) into~(\ref{13}), we have
 \begin{align}\label{17}
\begin{array}{l}
{p_{\rm{c}}}\left( {\theta ,r} \right) = \sum\limits_{i = 0}^M {\sum\limits_{j = 0}^{{m_i} - 1} {\sum\limits_{l = 1}^{{n_{{\rm{GHQ}}}}} {\frac{{{C_i}{a_l}{{\left( { - 1} \right)}^j}{W_{i,j}}\left( {{r^2}} \right)}}{{\exp \left( {\pi \lambda {r^2}{{\left( {{\theta  \mathord{\left/
 {\vphantom {\theta  {{\Omega _i}{b_l}}}} \right.
 \kern-\nulldelimiterspace} {{\Omega _i}{b_l}}}} \right)}^{{2 \mathord{\left/
 {\vphantom {2 \alpha }} \right.
 \kern-\nulldelimiterspace} \alpha }}}{E_{{h_0}}}} \right)}}} } } ,\\
{W_{i,j}}\left( t \right) = \sum\limits_{\left( {{N_1}, \cdots ,{N_j}} \right) \in {\mathbb{T}_j}} {\prod\limits_{q = 1}^j {\frac{{{{\left( {\frac{{2\pi \lambda t{E_{{h_q}}}}}{{{{\left( { - 1} \right)}^q}\alpha q!}}{{\left( {\frac{\theta }{{{\Omega _i}{b_l}}}} \right)}^{\frac{2}{\alpha }}}} \right)}^{{N_q}}}}}{{{N_q}!}}} }.
\end{array}
 \end{align}Note that if $j=0$, then ${W_{i,j}}\left( t \right) = 1$.
 Substituting the PDF ${f_R}\left( r \right) = 2\pi \lambda r\exp \left( { - \pi \lambda {r^2}} \right)$ and~(\ref{17}) into the coverage probability ${p_{\rm{c}}}\left( \theta  \right) = \mathbb{E}\left[ {{p_{\rm{c}}}\left( {\theta ,r} \right)} \right]$ and by some algebraic manipulations, we get
 \begin{align}\label{18}
{p_{\rm{c}}}\left( \theta  \right) = \sum\limits_{i = 0}^M {\sum\limits_{j = 0}^{{m_i} - 1} {\sum\limits_{l = 1}^{{n_{{\rm{GHQ}}}}} {\frac{{{C_i}{a_l}}}{{{{\left( { - 1} \right)}^j}}}\int_0^\infty  {\frac{{{W_{i,j}}\left( {{r \mathord{\left/
 {\vphantom {r {2\pi \lambda }}} \right.
 \kern-\nulldelimiterspace} {2\pi \lambda }}} \right)}}{{\exp \left( {r{G_i}} \right)}}dr} } } },
 \end{align}
 where ${G_i} = 1 + {\left( {{\theta  \mathord{\left/
 {\vphantom {\theta  {{\Omega _i}{b_l}}}} \right.
 \kern-\nulldelimiterspace} {{\Omega _i}{b_l}}}} \right)^{{2 \mathord{\left/
 {\vphantom {2 \alpha }} \right.
 \kern-\nulldelimiterspace} \alpha }}}{E_{{h_0}}}$. Because of the fact that ${W_{i,j}}\left( t \right) = \sum\limits_{\left( {{N_1}, \cdots ,{N_j}} \right) \in {\mathbb{T}_j}} {{A_j}{t^{\sum\nolimits_{q = 1}^j {{N_q}} }}}$, where $A_j$ is given in Theorem~\ref{Theorem:1}, the integral term of~(\ref{18}) can be rewritten as
 \begin{align}\label{19}
\begin{array}{l}
\int_0^\infty  {\exp \left( { - r{G_i}} \right){W_{i,j}}\left( {{r \mathord{\left/
 {\vphantom {r {2\pi \lambda }}} \right.
 \kern-\nulldelimiterspace} {2\pi \lambda }}} \right)dr} \\
 = \sum\limits_{\left( {{N_1}, \cdots ,{N_j}} \right) \in {\mathbb{T}_j}} {{A_j}\int_0^\infty  {\exp \left( { - r{G_i}} \right){r^{\sum\nolimits_{q = 1}^j {{N_q}} }}dr} }.
\end{array}
 \end{align}
 Finally, by evaluating the integral in~(\ref{19}) based on~\cite[Eq. (3.35.3)]{gradshteyn2007book} and with some algebraic manipulations in~(\ref{18}) yields~(\ref{9}). This completes the proof.
 \end{IEEEproof}
 Obviously,~(\ref{9}) avoids complicated integrals or high order differential calculations.
 As the popular fading distributions such as small-scale fading models, line-of-sight shadowing models and traditional composite fading/shadowing models are special cases of double shadowed fading~\cite[Fig. 1(b)]{chun2017twc}, the closed-form expression of the coverage probability obtained is generic. When the instantaneous shadowing of lognormal distribution is ignored, i.e. $\sigma_S=0$dB, Theorem~\ref{Theorem:1} is reduced to the $\kappa$-$\mu$ shadowed fading scenario experienced by the desired signal. Most importantly, the theoretical analysis of coverage probability concerning the interfering fading is avoided and it is replaced by simple calculations ${E_{{h_0}}}$ and ${E_{{h_q}}}$.


\vspace{-2mm}
\section{Numerical results}
\begin{figure}
\centering
\includegraphics[width=0.75\columnwidth]{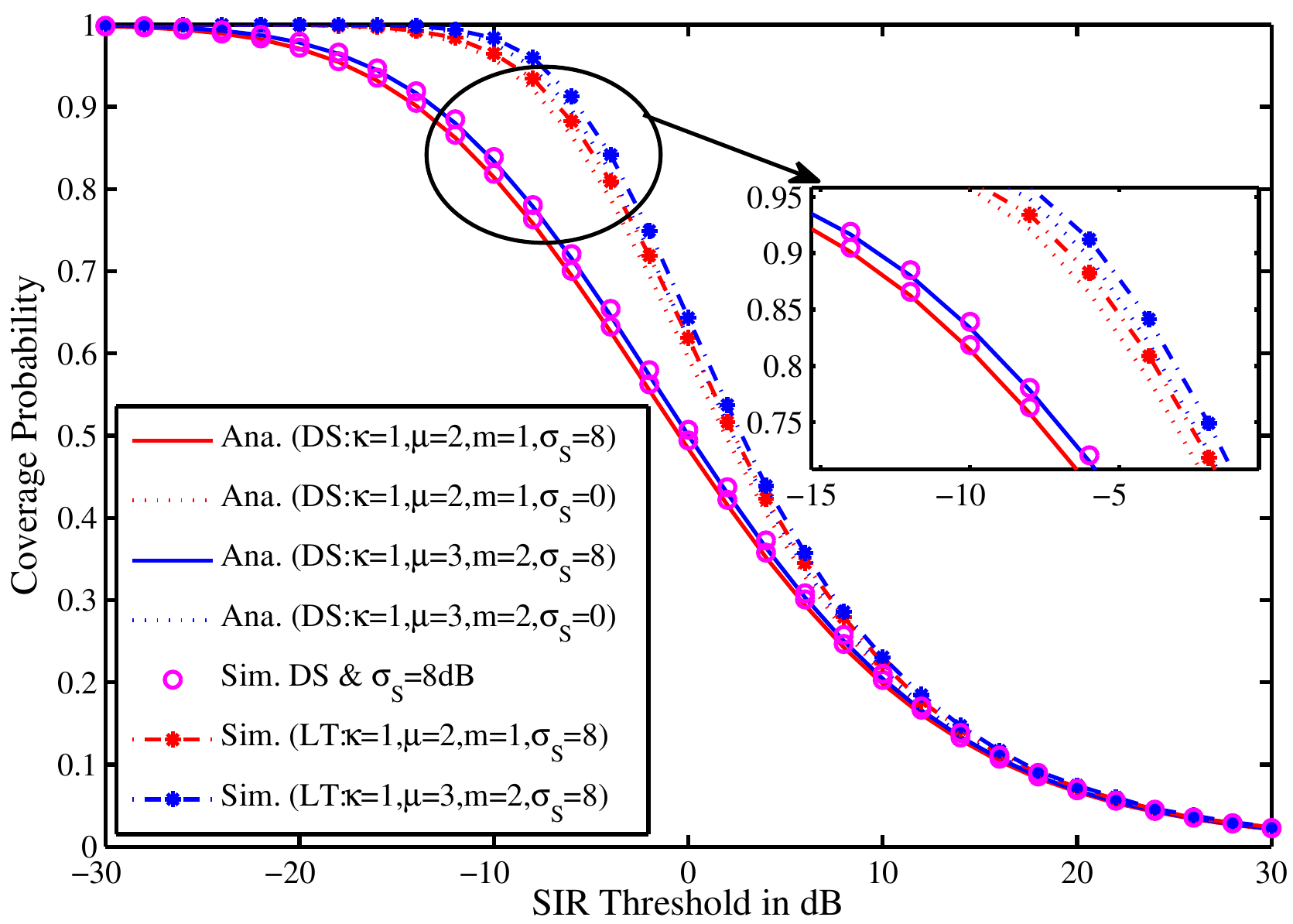}
\caption{Coverage probabilities with $DS$ and $LT$ for $M=\mu$.}
\label{fig:2}
\end{figure}

\begin{figure}
\centering
\includegraphics[width=0.75\columnwidth]{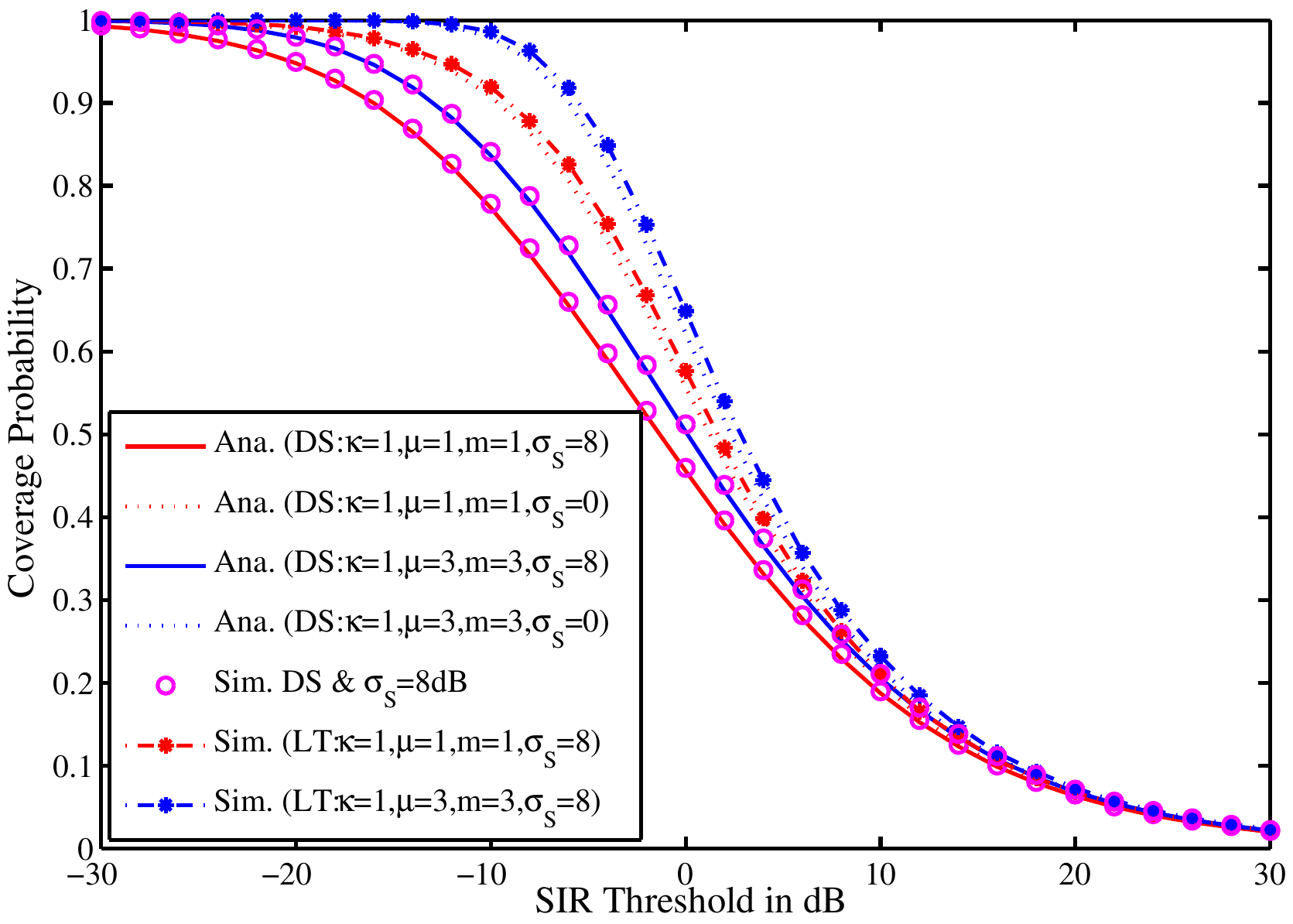}
\caption{Coverage probabilities with $DS$ and $LT$ for $M=m-\mu$.}
\label{fig:3}
\end{figure}

In this section, we validate the accuracy of derived coverage probability through Monte Carlo and make a comparison between the double shadowed fading and long-term shadowed scenario. For notational simplicity, we use $DS$ and $LT$ to stand for the double shadowed fading and comparative term of long-term fading respectively, which are experienced by the desired signal. The interfering signals experience Rayleigh/Lognormal fading and Rayleigh fading separately. The simulation parameters of the Poisson network are $\lambda  = {10^{ - 7}}$, $\alpha=4$ and $P=1$. For numeral calculations, we set ${n_{{\rm{GHQ}}}} = 32$. In Fig.~\ref{fig:2}, the coverage probability is plotted for the $DS$ model when $M=\mu$ (i.e. $\mu  > m$). It can be observed that the analytically obtained results closely match the simulation results. Intuitively, the coverage probabilities increase as parameters of $m$ and $\mu$ increase and decrease as the values of $\sigma_S$ increase. The coverage probability of the $LT$ model is also considered in Fig.~\ref{fig:2}. Comparing the coverage probabilities of the $DS$ model and $LT$ model, we can see that the instantaneous property of large scale fading makes a great difference in coverage. The $DS$ model provides a pessimistic coverage. This is due to the fact that choosing the serving BS depends solely on the distance. Hence, the instantaneous property of large scale fading is presented. When $M=m-\mu$ (i.e. $\mu  \leq m$), the coverage probabilities of the $DS$ model and $LT$ model are shown in Fig.~\ref{fig:3} and exact matches can be seen for various fading parameters. Similarly, it reveals the same conclusions as the Fig.~\ref{fig:2}.

\vspace{-2mm}
\section{Conclusion}
This paper analyzes the downlink coverage probability for a Poisson network when the desired signal experiences double shadowed fading and the interfering signals experience arbitrary fading. The nearest BS association policy without shadowing is considered. The exact and approximate expressions for the double shadowed fading composed of a lognormal distribution and $\kappa$-$\mu$ shadowed fading with integer fading parameters are provided. Based on the resulting expression, the generic and closed-form expression for the downlink coverage probability is successfully derived and verified through simulation. The obtained expression of the coverage probability indicates that the theoretical analysis of coverage probability about the interfering fading is avoided. In addition, compared to the long-term shadowed environment, numerical simulations show that the instantaneous shadowing of large scale fading is bad for the coverage of a Poisson network.

\vspace{-2mm}
\bibliographystyle{IEEEtran}

\end{document}